# Noise-Based Logic Hyperspace with the Superposition of $2^N$ States in a Single Wire

Laszlo B. Kish[a], Sunil Khatri, Swaminathan Sethuraman

Texas A&M University, Department of Electrical and Computer Engineering, College Station, TX 77843-3128, USA

**Abstract**. In the introductory paper, [Physics Letters A 373 (2009) 911–918], about noise-based logic, we showed how simple superpositions of single logic basis vectors can be achieved in a single wire. The superposition components were the *N* orthogonal logic basis vectors. Supposing that the different logic values have "on/off" states only, the resultant discrete superposition state represents a single number with *N* bit accuracy in a single wire, where *N* is the number of orthogonal logic vectors in the base. In the present paper, we show that the logic hyperspace (product) vectors defined in the introductory paper can be generalized to provide the discrete superposition of $2^N$ orthogonal system states. This is equivalent to a multi-valued logic system with $2^{2^N}$ logic values per wire. This is a similar situation to quantum informatics with *N* qubits, and hence we introduce the notion of noise-bit. This system has major differences compared to quantum informatics. The noise-based logic system is deterministic and each superposition element is instantly accessible with the high digital accuracy, via a real hardware parallelism, without decoherence and error correction, and without the requirement of repeating the logic operation many times to extract the probabilistic information. Moreover, the states in noise-based logic do not have to be normalized, and non-unitary operations can also be used. As an example, we introduce a string search algorithm which is $O(\sqrt{M})$ times faster than Grover's quantum algorithm (where *M* is the number of string entries), while it has the same hardware complexity class as the quantum algorithm.

---

[a] Until 1999: L.B. Kiss

## 1. Introduction: Noise-based logic

Recently a new type of deterministic logic scheme [1] was introduced, which utilized the orthogonality of independent stochastic processes (noise processes). Generally, for arbitrary independent reference noises $V_i(t)$ $(i = 1,...,N)$ :

$$\langle V_i(t)V_j(t) \rangle = \delta_{i,j} \qquad (1)$$

where $\delta_{i,j}$ is the Kronecker symbol (for $i = j$, $\delta_{i,j} = 1$, otherwise $\delta_{i,j} = 0$). Due to Equation 1, the $V_i(t)$ processes can be represented by orthogonal unit vectors in multidimensional space, thus we can use the term *logic basis vectors* for the reference noises, and introduce the notion of an $N$-*dimensional logic space*, with *logic state vectors* in it. "Deterministic logic" means here that the *idealistic* logic framework is completely *independent* of any notion of probability (there are no probabilistic rules), just like in Boolean logic. Of course, in a hardware realization, random bit errors may still occur due to non-idealistic components (similarly to binary logic), however the logic scheme stays deterministic (with errors).

By using this multidimensional space along with linear superposition, logic vectors and their superpositions can be defined, which results in a large number of different logic values, even with a relatively low number $N$ of basis vectors. For example, when using binary superposition coefficients that have only on/off possibilities of the reference noises, the number of possible logic values is $2^{2^N}$. It is important to emphasize that such a logic state vector is measured in a single wire [1], not on an ensemble of parallel wires. Thus using a linear binary superposition of basis vectors, $X(t) = \sum_{i=1}^{N} a_i V_i(t)$, where the $X(t)$ represents a single number with $N$ bit resolution, residing in a single wire. The coefficients $a_i$ are weights for each of the reference noise variables $V_i(t)$.

It was also shown that the noise-based logic scheme can replace classical digital gates, and the basic Boolean building elements (INVERTER, AND, OR, and XOR) were introduced in the noise-based scheme, in a deterministic way, with digital accuracy. The noise-based logic has several potential advantages, such as reduced error propagation and power dissipation, even though it may need larger numbers of circuit elements and binary operations may run somewhat slower.

In [1], another interesting property of the noise-based logic space was introduced that the product of two different (orthogonal) basis vectors is orthogonal to both the original noises. This property yields a logic hyperspace.

If $i \neq k$ and $H_{i,k}(t) \equiv V_i(t)V_k(t)$ then for all $n = 1...N$, $\langle H_{i,k}(t)V_n(t) \rangle = 0$. (2)

$H_{i,k}(t)$ is referred to as a hyperspace vector.

A similar operation can be done with the pairs of hyperspace vectors or with hyperspace vectors and basis vectors (with non-overlapping indexes), to grow the hyperspace [1], for example:

If $L_{i,k,l,m}(t) \equiv H_{i,k}(t)H_{l,m}(t)$ then $\langle L_{i,k,l,m}(t)V_n(t)\rangle = 0$, $\langle L_{i,k,l,m}(t)H_{p,q}(t)\rangle = 0$ (3)

If $L_{i,k,l}(t) \equiv H_{i,k}(t)V_l(t)$ then $\langle L_{i,k,l,m}(t)V_n(t)\rangle = 0$, $\langle L_{i,k,l,m}(t)H_{p,q}(t)\rangle = 0$ (4)

provided $i \neq k \neq l \neq m$. The same type of operations can be continued to generate new hyperspace elements, until we run out of non-overlapping groups of coordinate indexes from the original space. With $N$ basis vectors, a $2^N$ dimensional hyperspace can be generated in this manner.

This means that in the cases outlined above, the multiplication operation leads out of the original $N$-dimensional space and introduces higher, new dimensions that are orthogonal to each other and to the basis vectors of the original space.

In the present paper, we will utilize the hyperspace illustrated by Equations 2-4 [1] to introduce noise-based vectors that will be shown to represent a superposition of $2^N$ separate numbers in a single wire.

We will see that there are striking similarities, as well as key differences between this system and the logic states in quantum informatics, thus we will introduce a notation similar to that of quantum informatics to identify the similarities and differences more easily.

We will introduce the following simplifications:

*i.* We suppose idealistic noise-based elements (ideal multipliers, etc.) and leave the question of the practical implications of non-ideality for the subject of future studies.

*ii.* Similarly, here we do not discuss how to remove statistical inaccuracies and reach digital accuracy by threshold operations, but we refer to [1] for that.

*iii.* We use a real-number implementation of noise-based logic because it seems to be enough, however, we note here that complex numbers can also be represented by doubling the number of reference noises and related circuits.

*iv.* Here we focus on the multidimensional hyperspace and its information expressiveness and processing potential.

*v.* We do not deal with the general questions of operators. However, we introduce some simple operator examples, including operators for a noise-based string search engine, which will be used as a means to describe the potentials of noise-based logic.

## 2. Utilizing the logic hyperspace. The *noise-bit* and the number of states.

Similarly to quantum informatics where qubits are represented by, for example, the up/down states of a spin, for the noise-based hyperspace we introduce the *noise-bit*.

### *2.1 The noise-bit with two noises: the noise-based digital system with $2^N$ parallel states*

To demonstrate the analogy with the up and down spin states in quantum informatics, we can use two orthogonal noises (independent noise sources) $V_j^{(0)}(t)$ and $V_j^{(1)}(t)$ to represent the two alternative states, 0 and 1, of the *j*-th noise-bit, respectively (altogether $2N$ noise sources). To construct a basis vector element of the $2^N$ dimensional hyperspace, we create the product:

$$V_1^{(k_1)}(t)V_2^{(k_2)}(t)...V_N^{(k_N)}(t) = \prod_{j=1}^{N} V_j^{(k_j)}(t) , \qquad (5)$$

where the $k_j$ index is either 0 or 1 in accordance with the status of the *j*-th noise-bit. Because of the binary nature of the $k_j$ indices of this product, the dimension of this hyperscape is $2^N$. Each of these $2^N$ products is orthogonal to the others. For the sake of notational simplicity, in the rest of the paper, we omit the time variable (*t*) of noises because after computations are performed on the vectors, (*t*) will not exist in the results. The product given by Equation 5 is a basis element of the $2^N$ dimensional noise-based hyperspace. In this space, a $2^N$ dimensional vector *X* is a superposition of these base vectors. Generally,

$$X = \sum_{\substack{i=1 \\ k_{i,j}=0,1}}^{2^N} a_i \prod_{j=1}^{N} V_j^{(k_{i,j})} \qquad (6)$$

where the coefficients $a_i$ are real numbers. Such a vector *X* carries up to $2^N$ independent numbers in the form of the superposition coefficients $a_i$. Thus *N* noise-bits can carry up to $2^N$ independent numbers just like *N* qubits carry $2^N - 1$ numbers, where the -1 is due to the normalization condition of coefficient squares in quantum informatics; this condition is not necessary in noise-based logic (though it can be implemented).

We can follow the quantum notation and represent a $2^N$ dimensional state vector in (Equations 5,6) as:

$$V_1^{(k_1)}(t)V_2^{(k_2)}(t)...V_N^{(k_N)}(t) = \prod_{j=1}^{N}V_j^{(k_j)}(t) = |k_1,k_2,...k_N\rangle \tag{7}$$

As an example, the $2^N$ orthogonal basis elements of a 3 noise-bit system are as follows:

$$V_1^0V_2^0V_3^0 = |0,0,0\rangle, \quad V_1^1V_2^0V_3^0 = |1,0,0\rangle, \quad V_1^0V_2^1V_3^0 = |0,1,0\rangle, \quad V_1^0V_2^0V_3^1 = |0,0,1\rangle,$$
$$V_1^1V_2^1V_3^0 = |1,1,0\rangle, \quad V_1^0V_2^1V_3^1 = |0,1,1\rangle, \quad V_1^1V_2^0V_3^1 = |1,0,1\rangle, \quad V_1^1V_2^1V_3^1 = |1,1,1\rangle \tag{8}$$

## *2.2 The noise-bit with a single noise: the noise-based digital system with $2^N$ parallel states*

It may be more practical for most applications to use only a single noise for each noise-bit and then the two states of the given noise-bit will be represented by "noise", $|1\rangle$, or "vacuum" $|0\rangle$. In this case, the $|0\rangle$ states are the same for each noise-bit, that is they are "degenerated" from a quantum point of view, however from an informatics point of view, we still have $2^N$-1 orthogonal elements in the basis, thus the space is $2^N$-1 dimensional by using $N$ noise-bits and $N$ noises (instead of $2N$ noises as in Section 2.1). As an example, the $2^N - 1$ orthogonal basis elements of the "degenerated" 3 noise-bit system are as follows:

$$no\ signal = |0,0,0\rangle, \quad V_1 = |1,0,0\rangle, \quad V_2 = |0,1,0\rangle, \quad V_3 = |0,0,1\rangle,$$
$$V_1V_2 = |1,1,0\rangle, \quad V_1V_3 = |1,0,1\rangle, \quad V_2V_3 = |0,1,1\rangle, \quad V_1V_2V_3 = |1,1,1\rangle \tag{9}$$

In Figure 1, the practical realization of the basis described by Equation 9 is shown. The circles are the input points of the reference noises (noise-bits) the arrows within any line show the propagation of the noise signal, and the free-end arrows are the output points; each one provides a noise orthogonal to the noises at the other outputs. $2^N$-$N$-1 analog multipliers are needed in all.

From this point, to construct a discrete superposition of any of the $2^N$-1 basis vectors, only an analog adder is needed for a superposition with binary coefficients, as shown in Figure 2. The superposition will reside on a single wire, and all of its components are continuously accessible in a parallel manner without the need of repeated measurements (because of the absence of collapse-of-the-wavefuntion type effects like in quantum informatics). This is a real, classical parallelism. The price is that the circuit complexity for general-purpose superpositions scales as $2^N$. However, for certain special purpose applications, similar tricks can be applied as in the initialization operations of the Shor-algorithm in quantum computing, which makes a uniform superposition of $2^N$-1 numbers possible with $O(N)$ circuit complexity, as described later in the text.

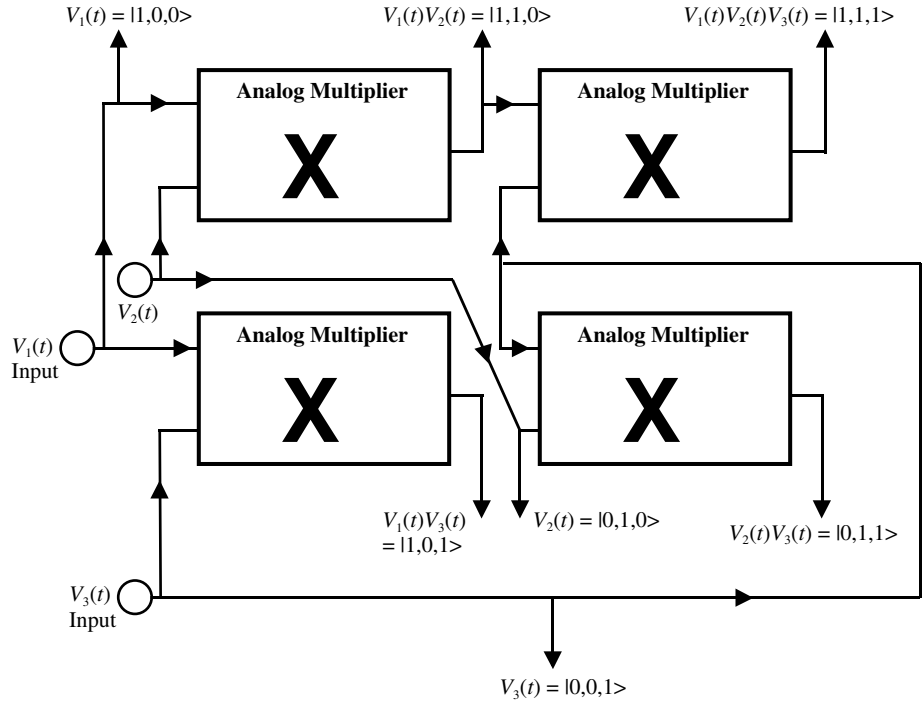

**Figure 1.** Making the system of $2^N-1$ orthogonal noise-based binary vector states comprising the hyperspace vectors with $N$=3.

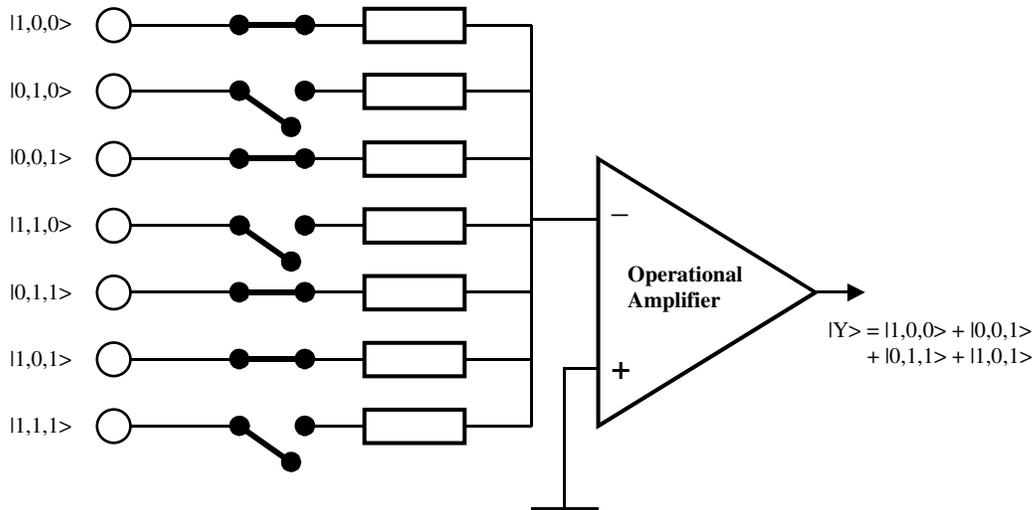

**Figure 2.** Example of controlled adder for discrete superpositions made by an operational amplifier. All resistors have the same value.

## 2.3 Measurement of the state

The detection of a certain hyperspace basis state in the input superposition $X(t)$ is done in the same manner as it is described in [1]: by multiplying the signal with the relevant basis state and then executing a time average operation. Using Dirac's angle-bracket notation of scalar product, quantum physics provides a convenient notation:

$$\langle X(t)V_i(t)\rangle = \langle X|V_i\rangle \quad , \tag{10}$$

where the left-hand side of the equation shows the actual multiplication and time average operations (cross-correlator); the right-hand side shows the same with the brackets; $X(t) \equiv |X\rangle$; and $V_i(t) \equiv |V_i\rangle$.

Equation 10 produces a number, non-zero or zero (a DC voltage in practice), depending on whether the $|X\rangle$ superposition has the $|V_i\rangle$ component in it, or not. In a noise-based logic gate, the output of the gate must be an element of the noise hyperspace (generally a superposition), thus this DC voltage is typically driving analog switches (followers or inverters) that connect the noise-based logic vectors to the output. This operation is represented by a product of the DC voltage and the appropriate noise-based logic vector [1]. For arguments and estimations on why the analog switches can be of low-quality (high switching error rate) and of low-energy dissipation, and yet yield a logic operation with high accuracy (low error rate), see [1].

As an illustration, let us detect if the $V_1^1 V_2^0 V_3^1 = |1,0,1\rangle$ component is present in the superposition $aV_1^1 V_2^0 V_3^1 + bV_1^0 V_2^1 V_3^0 = a|1,0,1\rangle + b|0,1,0\rangle = |Y\rangle$. Let us suppose that the $|H\rangle$ noise vector stands for the "*True*" (yes) answer and the $|L\rangle$ (orthogonal noise to $|H\rangle$) stands for "False" (no).

$$|Y\rangle = \overline{\langle X|1,0,1\rangle}\,|L\rangle + \langle X|1,0,1\rangle\,|H\rangle \tag{11}$$

where the overbar represents the inverting operation (i.e. switching the zero DC value of the scalar product into 1 and vice versa, which may not have a simple quantum analogy). The simple realization of this function with one multiplier, one time average unit and two analog switches (inverter and follower) can be seen in Figure 3.

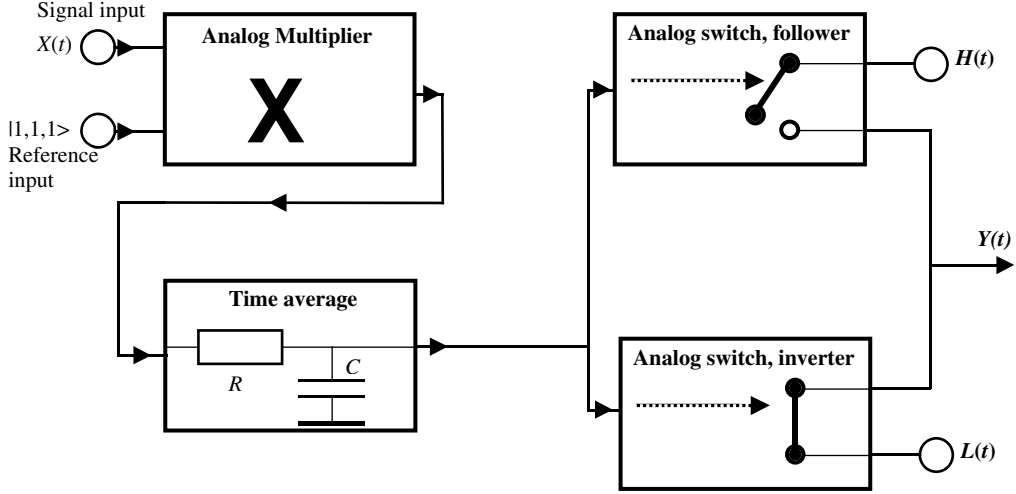

**Figure 3.** Detecting the component |1,1,1> in the superposition $X(t)$. All the other orthogonal elements need a similar circuitry for detection. $L(t)$ represents the logic value (noise) for "False" and $H(t)$ the logic value for "True".

It is very important to recognize that the measurement result of the unit described by Equation 11 and Figure 3 is deterministic, not probabilistic. By using such a single unit and probing a hyperscape superposition by $2^N$ subsequent measurements, we have a total and accurate description of the system state. On the other hand, by using $2^N$ units like that in Figure 3, the system state can be probed by a single measurement. It is equivalent to exporting $2^N$ words, each of length $N$, via an $N$-bit serial port to output the whole information content of the system. We will see in Section 3 that quantum information systems, which have exactly the same information content, also act like a serial port, however the corresponding quantum system is a probabilistic serial port that does not guarantee a full description of the system state after $2^N$ measurements. The quantum system require many more measurements to extract this information, in a probabilistic form, with limited accuracy.

As mentioned above, in the noise-based logic system, the unit shown in Figure 3 can be replicated $2^N$ times, where each unit is driven by a different component of the $2^N$ dimensional hyperspace basis. This is equivalent to a large parallel port that is able to extract of the total information from the system in a single clock step. Such a speedup is impossible in a quantum system because, due to the collapse of the wavefunction, it always behaves as a probabilistic serial port.

Another relevant example is the noise-based logic projection operator $P^{|1,0,1\rangle}$ for the component $|1,0,1\rangle$ of the input superposition $|X\rangle$ is shown in Fig.4. The output, which is

$$|Y\rangle = P^{|1,0,1\rangle}|X\rangle \qquad (12)$$

will produce $|1,0,1\rangle$ whenever the $|1,0,1\rangle$ element is present in the superposition $|X\rangle$,

otherwise it produces zero. In effect, this queries the membership of $|1,0,1\rangle$ in $|X\rangle$. In practical applications (see Figure 4) the threshold voltage of the switch must be properly selected (or an amplifier stage must be introduced before the switch) if the amplitude of the superposition component is too small.

The full projection operator $P^{2^N}$ consists of $2^N$ instances of this operator with joined signal inputs and joined outputs, where the reference input of each instance is fed by different element of the hyperspace $|0,0,0\rangle,...,|1,1,1\rangle$.

$$P^{2^N} = \sum_{i=1}^{2^N} P^{|k_{1,i},k_{2,i},...k_{N,i}\rangle} \quad , \tag{13}$$

where $k_{j,i}$ = 0, or 1. The operator $P^{2^N}$ will produce, at the output, the superposition of those hyperspace elements that are present in the input superposition. In effect, this full projection operator is a buffer or repeater unit.

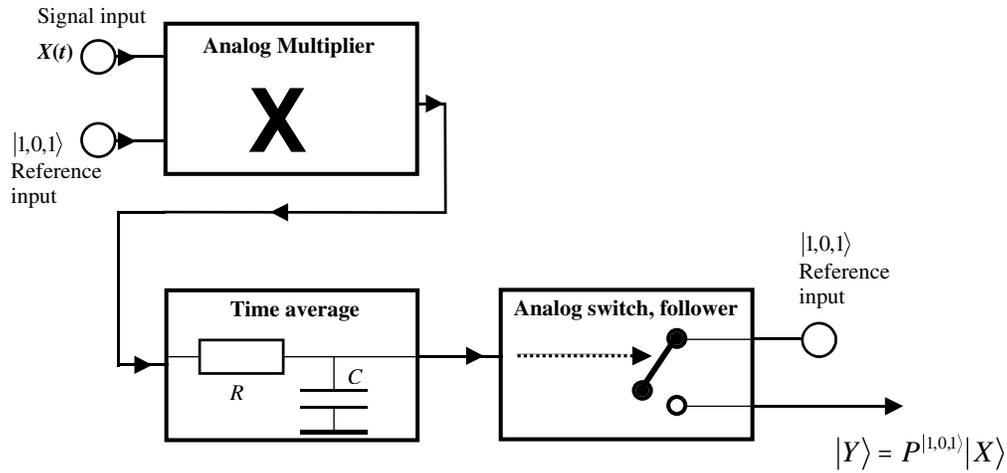

**Figure 4.** Noise-based logic projection operator $P^{|1,0,1>}$ for the component $|1,0,1>$ in the input superposition $X(t)$.

### 2.4 Noise-based hardware and its complexity

In general-purpose applications, to realize the full power of noise-based logic, the hyperspace must be generated and each superposition coefficients must be controllable and accessible separately. This task naturally has an $O(2^N)$ hardware complexity just like general-purpose quantum computing applications do [5,6]. However, *special-purpose applications*, similar to quantum computers, may have the potential to be realized with low complexity.

For example, similar to the initialization of Shor's algorithm in quantum computing, to set up the uniform superposition $|0,0,...,0\rangle + ... + |1,1,...,1\rangle$ of the $2^N$ different hyperspace vectors (that is a superposition of $2^N$ of different digital numbers with *N* bit resolution), we can use the following algorithm that has an $O(2^N)$ complexity. The uniform superposition can be achieved with *N* additions and *N* multiplications:

$$|Y\rangle = \left[V_1^{(0)} + V_1^{(1)}\right]\left[V_2^{(0)} + V_2^{(1)}\right]...\left[V_N^{(0)} + V_N^{(1)}\right] = \prod_{j=1}^{N}\left[V_j^{(0)} + V_j^{(1)}\right] \quad (14)$$

or with the bracket notation,

$$|Y\rangle = |0,0,...,0\rangle + ... + |1,1,...,1\rangle =$$
$$= \left[|0\rangle_1 + |1\rangle_1\right]\left[|0\rangle_2 + |1\rangle_2\right]...\left[|0\rangle_N + |1\rangle_N\right] = \prod_{j=1}^{N}\left[|0\rangle_j + |1\rangle_j\right] \quad (15)$$

This transformation is similar to the Achilles' Heel problem in logic. Let $B = \{0,1\}$ and $x_i^k \in B$, $i = 1,...,N$ and $k = 0,1$. For a logic function $F$ expressed in Conjunctive Normal Form (CNF) as $F = \prod_{i=1}^{N}(x_i^0 + x_i^1)$, the corresponding Disjunctive Normal Form (DNF) expression of $F$ is $\sum_{k_j \in 0,1} x_1^{k_1} x_2^{k_2}...x_N^{k_N}$. The complexity of the DNF of $F$ is $O(2^N)$ while the complexity of the CNF is $O(N)$, only.

## 3. On the similarities with and differences from quantum computers

In this section, we summarize the similarities with and differences from quantum computers as pointed out above, with additional comments. The major differences of noise-based algorithms compared to quantum informatics are that noise-based logic is deterministic and each superposition element is instantly accessible with the usual high (digital) accuracy (low error probability), via a real hardware parallelism, without decoherence, without error correction, and without the need of repeating the logic operation many times to extract the probabilistic information. These are all problems in quantum informatics. Moreover, the states do not have to be normalized and non-unitary operations can also be used.

### *3.1 General-purpose computers*

Both noise-based and quantum computers have a hardware complexity of $O(2^N)$ where *N* is the number of noise-bits or qubits, respectively. The noise-based computer operates in a deterministic manner, where all the respective hardware units produce information, while the quantum computer works in a probabilistic fashion where a fraction of the respective hardware units produce information during a single run. Thus general-purpose quantum computers will generally require a significantly longer time to solve the same problem of arbitrary nature.

*3.2 Special-purpose computers*

Special-purpose computers (similarly to an analog computer) can solve only one problem or a narrow class of problems, however they can do this faster and with fewer resources. Here the comparison between noise-based and quantum computers must be done in a case-by-case manner. For example, in the next section, we will show that the noise-based string search engine outperforms Grover's quantum search engine with the same hardware complexity.

Note that theoretically it is possible that the superposition of the full set of $2^N$ hyperspace elements may be needed to solve a problem, but the input and output hardware complexity to generate and evaluate the result does not require to be $2^N$. One example is given in Equations 14, 15, where the superposition can be generated with an input operation, which has $O(N)$ complexity, similarly to Shor's quantum algorithm. This holds for both the noise-based and the quantum algorithm. The rest of the comparison about the Shor-algorithm has not been carried out yet, so it is currently unclear if it can be done with hardware complexity polynomial in $N$.

Similarly, using sampling, multiplexing, nonlinear operations (such as logarithmic and exponential amplifiers) may help to generate special-purpose superpositions in noise-based computers, and some of these options seem to be available also in quantum computers as well.

**4. Faster string search algorithm than Grover's quantum search algorithm**

Finally, as a simple example of the potential quantum-related applications, we show a simple string search engine with noise-based logic, that has the same hardware complexity class as the related Grover's quantum search engine [2]. However its speed is $O(1)$; compared with the $O(1/\sqrt{M})$ speed of Grover's quantum algorithm, where $M$ is the number of string entries. Consider the string register $\{S_j\}$ and its corresponding address register $\{A_j\}$, each with $N$ entries.

$$\left|S_j\right\rangle = \left|V_1^{(k_{1,j})}, V_2^{(k_{2,j})}, ..., V_N^{(k_{N,j})}\right\rangle \tag{16}$$

$$\left|A_j\right\rangle = \left|W_1^{(r_{1,j})}, W_2^{(r_{2,j})}, ..., W_N^{(r_{N,j})}\right\rangle \tag{17}$$

where $V_i$ is the $i$-th noise-bit in the string, $W_i$ is the $i$-th noise-bit in the address, $k_{i,j}$ represents the state (0 or 1) of the $i$-th noise-bit in the $j$-th string and $r_{i,j}$ represents the state (0 or 1) of the $i$-th noise-bit in the address of the $j$-th string.

Consider register $\{R_j\}$ with $2^N$ elements (line vectors) that are 2N noise-bits long. The j-th line of $\{R_j\}$ is the j-th string $S_j$ and its address string $A_j$ combined to form a single line vector with lengths 2N noise-bits:

$$|R_j\rangle = |A_j\rangle|S_j\rangle \tag{18}$$

In this way, the $\{A_j\}$ and $\{S_j\}$ registers are "entangled" while forming $\{R_j\}$. This operation is hardwired in the architecture of the string-searching engine. This provides $\{R_j\}$ at the same time when the strings are loaded into the register $\{S_j\}$. Similarly to quantum computers, this can be done in a single step if the sufficient number of parallel inputs is available. After the strings are uploaded, the string search algorithm can be executed. The algorithm is a single computing step described by the following operation:

$$P_A^{2^N}\left[|S_i\rangle\sum_{j=1}^{2^N}|R_j\rangle\right] = P_A^{2^N}\left[\langle S_i|S_i\rangle A_i + |S_i\rangle\sum_{\substack{j=1\\j\neq i}}^{2^N}|R_j\rangle\right] = |A_i\rangle , \tag{19}$$

where the full projection operator $P_A^{2^N}$ is working over the A register, that is, it can project any of the $2^N$ dimensional base vectors (the string addresses) from the A hyperspace. After the projection operator $P_A^{2^N}$ acts on the product of the i-th string $|S_i\rangle$ and the superposition of the elements in R, the operation over the second term in the intermediate section of Equation 19 yields zero, and the first term there will result in the address $|A_i\rangle$. In this way, if a string $|S_i\rangle$ is presented to the noise-based search engine, it returns $|A_j\rangle$. It can therefore perform the operation, which is the equivalent of a phonebook lookup, in 1 step. Note, it can also do a reverse lookup by replacing $|S_i\rangle$ by $|A_j\rangle$ in the left-hand-side of Equation 19. Note also, that Equation 19 can also implement the functionality of a cache memory.

Building this noise-based search engine for M different strings ($M = 2^N$ at most) requires an $O(M)$ hardware complexity. This is the same situation as with Grover's quantum search engine because there the M strings must be inputted at the same time, in a parallel way, from a classical physical system. That requires M parallel control units working at the input of a quantum computer (which is equivalent to an $O(M)$ hardware complexity).

It has been shown [3] that Grover's algorithm is optimal, thus no faster quantum algorithm can be proposed for this goal. However, Grover's algorithm was originally posed to solve a restricted version of the arbitrary string search algorithm shown above. Equation 19 can be used to perform Grover's search [2] in $O(1)$ time as follows. In Grover's search, we have M bits, of which one bit has a different value than the others.

Grover's algorithm finds the errant bit in $O(\sqrt{M})$ steps. In noise-based logic, this can be achieved by setting

$$|S_j\rangle = V_1^{(k_1)}, \; j = 1,2,...,i-1, i+1,...,N, \text{ and } S_i = V_1^{(\bar{k}_1)} \qquad (20)$$

Here $k_1 \in \{0,1\}$ and $\bar{k}_1$ is the complement of $k_1$. Also $|A_j\rangle$ is assigned to the address (binary representation) of $j$. For example, suppose $N = 5$ and $j = 10$, and $|A_{10}\rangle = |w_1^0 w_2^1 w_3^0 w_4^1 w_5^0\rangle$. Now, we simply apply Equation 19 and the output of the noise-based computer provides the address of the errant entry (obtained in 1 step).

A variant of Grover's algorithm is where the searched string contains the superposition of $2^N - 1$ hyperspace elements. The algorithm needs to find the missing hyperspace element. This is semantically identical to Grover's algorithm that finds the single bit, which has a different value than other bits in a set. In this variant, the missing hyperspace element is the "bit" that has a different value than the other "bits" (which are hyperspace elements that are present in a superposition).

To implement this in noise-based logic, we use $M$ copies of the circuit in Figure 4, with joined outputs. Each copy is presented an input which is a superposition of the $M - 1$ hyperspace elements corresponding to the input of Grover's algorithm. Each copy of the circuit on Figure 4 is modified such that the follower switch is replaced by an inverter switch. Now, the circuit outputs the address of the $M$-th (errant) missing hyperspace element.

Note, that the circuit of Figure 4 allows us to realize true multivalued logic [10]. The reason why multivalued logic has not received much acceptance in VLSI design is that the supply voltage of VLSI circuitry (currently 1-2 Volts) does not allow for a reliable resolution of greater than 2 (i.e. binary) values. A multivalued function $f$ is a mapping $f : \prod_{i=1}^{S} B_i \to \prod_{j=1}^{T} C_j$, where $|B_i|, |C_j| > 2$. In the circuit of Figure 4, a multi-valued function can be implemented (for $S = T = 1$) by selecting the reference inputs on the left of Figure 4 and encoding the element of $B_i$ into the noise-based hyperspace. Similarly, the output is obtained by encoding the corresponding element of $C_j$ into the noise hyperspace. By joining several such outputs, we can implement $f : B_1 \to C_1$. This can be generalized to multiple inputs.

We have shown above that the speed of the noise-based algorithm is $O(\sqrt{M})$ times faster than the speed of Grover's quantum algorithm. The reason for this the same one mentioned above: the noise-based logic is free of the collapse of wavefunction and the probabilistic nature at quantum measurements, while it can also produce the same type of superpositions as quantum systems. In Grover's algorithm, the runs must be repeated to extract the result with sufficient accuracy [2-4] because of the probabilistic nature of quantum logic; however, in the noise-based algorithm no repetition is necessary.

It is important to note that the above special-purpose string-searching engine with noise-based logic is not meant to replace classical computers, just like Grover's algorithm is not meant to do that. The reason is the great hardware complexity in both systems. If noise-based logic will ever be used for search, it will likely be in a general-purpose computer, where classical binary gates are replaced by their noise-based binary equivalents [1] or by their multi-valued versions.

Finally, we give a rough (pessimistic) estimation for a silicon microprocessor chip of today. As of today, the number of transistors in a microprocessor chip is about 1 billion [11]. Let us suppose that each string vector has $N = 25$ elements and the number of elements in the address vectors is the same. If each element needs 10 transistors, then the required number of transistors for strings is about 336 million and the same number of transistors is used for the addresses, too. We can pessimistically suppose that the same number of transistors will be needed for the utility circuitry running the process. Thus, we end with the number 1 billion relevant today. The system will be able to handle a phonebook with $2^N$ = 33.5 million strings, each one 25 bit long. Supposing 200 GHz small signal bandwidth and a 500-fold slowdown [1] due to averaging the correlator outputs for digital accuracy, the clock frequency will be 0.4 GHz. Pessimistically supposing that, instead of a single operation, 4 operations will be needed to get the searched string address (due to resetting, etc), the total search time will be 10 nanoseconds. This is many orders of magnitude faster then any digital string search algorithm of today.

**5. Brain logic circuitry ?**

In paper [1], it is mentioned that the stochasticity of brain signals is inspired the noise-based logic system and that includes also the hyperspace introduced there [1]. Thus, we can naturally ask whether the multivalued hyperspace schemes developed in the present paper can be relevant for the neural circuitry of the brain. This question is studied and relevant possibilities are explored in a new paper submitted for publication [12]. The preliminary answer is yes: the same kind of hyperspace can be built with neurons and their spiky, stochastic signals (see [12]). The result in [12] have the potential to open new perspectives to explain information processing in the brain and the most important remaining question is whether those neural circuit elements [12] can indeed be found in the brain.

**6. Conclusions**

We show how noise-based logic [1] can be extended to operate on a noise hyperspace. With this enhancement, we are able to perform many computations with improved runtime and similar hardware complexity, compared to quantum computing. We demonstrate that Grover's algorithm can be done in $O(1)$ time and $O(M)$ circuit resources, compared to the $O(\sqrt{M})$ time and $O(M)$ resources for quantum logic. Also, noise-based logic elegantly allows us to implement multivalued logic [10], membership lookup in databases, phonebook (forward as well as reverse) lookups, caches and logic gates. This is achieved without wavefunction collapse issues (commonly experienced in quantum computing), yielding high speed and accuracy.

**Acknowledgements**

The authors are grateful to Suhail Zubairy for valuable discussions about the Grover's algorithm. LBK appreciates useful discussions with Julio Gea-Banacloche about quantum informatics.